\documentclass[a4paper,12pt]{article}
\usepackage[cp1251]{inputenc}
\usepackage[dvips,dvipdf]{graphicx,graphics}
\usepackage[english]{babel}
\usepackage{amsfonts,amssymb, amsmath, amsthm}
\usepackage{multicol}

\title{Exact solution of the Percus-Yevick integral equation for \textquotedblleft collapsing\textquotedblright\,hard spheres}

\date{}

\author{I.\,Klebanov, P.\,Gritsay, N.\,Ginchitskii}

\begin{document}
\maketitle

\begin{quotation}
\footnotesize
\begin{center}
\itshape
Chelyabinsk State Pedagogical University, Departament of Mathematics, 454080 Chelyabinsk, Russia
\end{center}

By Wertheim-method the exact solution of the Percus-Yevick integral equation for a system of particles with the \textquotedblleft repulsive step potential\textquotedblright\,interacting (\textquotedblleft collapsing\textquotedblright\,hard spheres) is obtained. On the basis of this solution the state equation for the \textquotedblleft repulsive step potential\textquotedblright\,is built and determined, that the Percus-Yevick equation does not show phase transition for \textquotedblleft collapsing\textquotedblright\,hard spheres.\\

PACS: 61.20.Ja,64.70.Ja
\end{quotation}

\begin{multicols}{2}
In 1963 Wertheim and Thiele independently gained the exact analytical solution of the Percus-Yevick integral equation for hard spheres \cite{Wertheim:63}-\cite{Thiele:63}. This solution is so far the only strict analytical one of non-linear integral equation for pair distribution function. The given work shows the possibility of Wertheim method to express the solution of the Percus-Yevick equation in closed analytical form for more complicated particles interaction potential --- repulsive step potential \cite{Ryzhov:03}\\
\begin{equation}
V(r)=
	\begin{cases}
		\infty, &r<a\\
		V_{0},&a\leq r\leq b\\
		0,&r>b.     
	\end{cases}
\end{equation}
where $V_{0}$ --- positive constant, $r$ --- particles range (\textquotedblleft collapsing\textquotedblright\,hard spheres).

Such a potential used to have a great application in modeling phase transition in liquid under the high pressure, isostructural phase transitions in crystals, transformations in colloid systems, etc. by means of molecular dynamics within the framework of thermodynamic perturbation theory \cite{Ryzhov:03,Mikheenkov:04}.\\
Let's consider the Percus-Yevick equation
\begin{equation}
\begin{split}
n_{2}(r)=1-n&\!\int\left[e^{\beta V(\vec{s})}-1\right]n_{2}(\vec{s})\times\\
&\times\left[n_{2}(\vec{r}-\vec{s})-1\right]\vec{ds}
\end{split}
\end{equation}
where $n_{2}(r)$ --- the pair distribution function, $\beta\!\!=\!\!\dfrac{1}{kT}$, $n$ --- the particle density.
Moving to bipolar coordinates and integrating for angle variable for the \textquotedblleft repulsive step potential\textquotedblright\,
we gain that

\begin{equation}
\begin{split}
&h(r)=Ar-2\pi n\int\limits^{a}_{0}h(s)ds\int\limits^{r+s}_{\left|r-s\right|}h(t)e^{-\beta V(t)}dt\,-\\
&-2\pi n(1-e^{-\beta V_{0}})\int\limits^{b}_{a}h(s)ds\int\limits^{r+s}_{\left|r-s\right|}h(t)e^{-\beta V(t)}\,dt
\end{split}
\end{equation}
where 
\begin{equation}
\begin{split}
h(r)=rn_{2}(r)&e^{\beta V(r)}=\\=&
\begin{cases}
-rC(r),&r<a\\
\dfrac{-rC(r)}{1-e^{-\beta V_{0}}},&a\leq r\leq b\\
n_{2}(r),&r>b
\end{cases}
\end{split}
\end{equation}
$C(r)$ --- direct correlation function. In the approach of Percus-Yevick
\begin{equation*}
C(r)=(1-e^{\beta V(r)})n_{2}(r)
\end{equation*} 
$n_{2}(r)=0$ at $r<a$, $C(r)=0$ at $r>b$, and $e^{-\beta V(t)}=e^{-\beta V_{0}}\Theta(t-a)\Theta(b-t)+\Theta(t-b)$; $\Theta(x)$ --- Haviside step function;
\begin{equation*}
\begin{split}
A=1+4\pi &n\int\limits^{a}_{0}\!\!s\,h(s)ds+\\+
&4\pi n(1-e^{-\beta V_{0}})\int\limits^{b}_{a}\!\!s\,h(s)ds 
\end{split}
\end{equation*} 
We take the one-side Laplace transform for (3) $\hat L(h(r))=\int\limits^{\infty}_{0}h(r)e^{-zr}dr$ and change the order of integration for $r$ and $t$, we finally obtain:
\begin{equation}
\psi(s)=\frac{\dfrac{A+\gamma z\delta(z)}{z^{2}}-L(z)}{1-\dfrac{2\pi n}{z}\left[L(z)-L(-z)\right]}
\end{equation}
where
\begin{equation*}
\begin{split}
&\psi(z)=\hat L(rn_{2}(r))=G(z)+e^{-\beta V_{0}}K(z)\\
&L(z)=\hat L(-rC(r))=F(z)+(1-e^{-\beta V_{0}})K(z)\\
&F(z)=\int\limits^{a}_{0}h(s)e^{-zs}ds\\
&K(z)=\int\limits^{b}_{a}h(s)e^{-zs}ds\\
&G(z)=\int\limits^{\infty}_{b}h(s)e^{-zs}ds\\
&\delta(z)=\alpha(z)-\alpha(-z)\\
&\gamma=2\pi\,ne^{-\beta V_{0}}(1-e^{-\beta V_{0}})
\end{split}
\end{equation*}
For further investigation we lead in the following function
\begin{equation}
H(z)=z^{4}\psi(z)\left[\frac{A+\gamma z\delta(z)}{z^{2}}-L(-z)\right]
\end{equation} 
Discussions like in \cite{Wertheim:64} show that
\begin{equation*}
H(z)=\lambda_{1}+\lambda_{2}z^{2}, 
\end{equation*}
where $\lambda_{1},\lambda_{2}$
--- are contstants. Not including $\psi(z)L(-z)$ from (5) and (6) and turning the Laplace transform into the $r\leq b$ area, we gain the explicit expression for $h(r)$
\begin{equation}
h(r)=-(C_{0}+C_{1}r+C_{2}r^{2}+C_{3}r^{4})
\end{equation}
where
\begin{equation*}
\begin{split}
&C_{0}=2\pi ne^{-\beta V_{0}}(\lambda_{1}k_{2}+k_{0}(\gamma\delta_{1}-l_{0}))\\
&C_{1}=\lambda_{1}(-1+2\pi ne^{-\beta V_{0}}k_{1})\\
&C_{2}=\pi n(-\lambda_{2}+\lambda_{1}k_{0}e^{-\beta V_{0}})\\
&C_{3}=-\frac{\pi n\lambda_{1}}{12}
\end{split}
\end{equation*}
and the constants $\lambda_{1},\lambda_{2},k_{0},k_{1},k_{2},l_{0}$ and $\delta_{1}$ as density, temperature and potential parametres $V_{0},a,b$ functions can be obtained from the system of equations
\begin{equation}
\begin{split}
&\lambda_{1}=A=1+4\pi n\int\limits^{a}_{0}rh(r)dr-4\pi n(1-e^{-\beta V_{0}})k_{1}\\
&\lambda_{2}=2\gamma\delta_{1}-2l_{0}-\frac{2\pi n}{3}\times\\
&\times\left[\int\limits^{a}_{0}r^{3}h(r)dr+(1-e^{-\beta V_{0}})\int\limits^{b}_{a}r^{3}h(r)dr\right]\\
&k_{m}=\frac{(-1)^{m}}{m!}\int\limits^{b}_{a}r^{m}h(r)dr,\,\,\,m=0,1,2,\\
&l_{0}=\int\limits^{a}_{0}h(r)dr+(1-e^{-\beta V_{0}})k_{0}\\
&\delta_{1}=\int\limits^{b}_{a}\int\limits^{b}_{s}h(s)h(t)(t-s)dtds
\end{split}
\end{equation}
Inserting (7) into (5), we gain the Laplace image form for $rn_{2}(r)$.\\
The system of \textquotedblleft collapsing\textquotedblright\, hard spheres state equation can be shown as
\begin{equation}
\begin{split}
&\frac{P}{nkT}=1-\frac{n}{6kT}\int \!\!rn_{2}(r)\left(\frac{dV}{dr}\right)\vec{dr}=\\
&=1+\frac{2\pi n}{3}\left[e^{-\beta V_{0}}a^{3}\tau(a)+(1-e^{-\beta V_{0}})b^{3}\tau(b)\right]
\end{split}
\end{equation}
where $\tau(r)=h(r)r^{-1}$ and inverse isothermic compressibility
\begin{equation}
\begin{split}
&\left(\frac{\partial P}{\partial n}\right)_{T}\frac{1}{kT}=\\
&=1-n\int \!\!C(r)\vec{dr}=\lambda_{1}(n,T)
\end{split}
\end{equation}
It can be seen from (10) and (8) that if $V_{0}>0$
\begin{equation}
\left(\frac{\partial P}{\partial n}\right)_{T}>0
\end{equation}
i.e. the Van der Waals loop is absent in the isotherm. This result coincides with those ones wich were taken in numerical analysis of the state equation (9).\\
Thus, the Percus-Yevick equation for the system of \textquotedblleft collapsing\textquotedblright \,hard spheres allow the solution in closed analytical form going into the Wertheim-Thiele classical solution for hard spheres when $a=b$. As in the case of hard spheres the Percus-Yevick solution doesnt show the phase transition in the system of \textquotedblleft collapsing\textquotedblright \,hard spheres.

\end{multicols}
\end{document}